\documentclass[a4paper,10pt]{article}

\usepackage{epsfig,graphicx}
\usepackage{amsmath,amssymb}

\begin{document}

\title{\bf High Frequency Dynamics and Third Cumulant of
Quantum Noise}
\author{ J. Gabelli and B. Reulet \\
         Laboratoire de Physique des Solides,\\
          UMR8502 b\^atiment
510, Universit\'e Paris-Sud 91405 ORSAY Cedex, France}
\date{}
\maketitle

\bigskip

Physics of current fluctuations has proven, during the last 15
years, to be a very profound topic of electron transport in
mesoscopic conductors (for a review, see ref. \cite{Blanter}).
Usually, current fluctuations are characterized by their spectral
density $S_2(\omega)$ measured at frequency $\omega$:

\begin{equation}\label{eq:S2}
 S_2(\omega) = \langle i(\omega) \, i(- \omega )\rangle,
\end{equation}
\vskip 0.3cm

\noindent where $i(\omega)$ is the Fourier component of the
classical fluctuating current at frequency $\omega$ and the
brackets $\langle . \rangle$ denote time averaging. In the limit
where the current can be considered as carried by individual,
uncorrelated electrons of charge $e$ crossing the sample (as in a
tunnel junction), $S_2(\omega)$ is given by the Poisson value
$S_2(\omega)=e\, I$ and is independent of the measurement
frequency $\omega$.  At sufficiently high frequencies, however,
this relation  breaks down and should reveal information about
energy scales of the system. In particular, in the quantum regime
$\hbar \omega > eV$ ($V$ is the voltage across the conductor), it
turns out that the noise cannot be seen as a charge counting
statistics problem anymore even for a conductor without intrinsic
energy scale. In this regime, the noise spectral density reduces
to its equilibrium value determined, at zero temperature, by the
zero-point fluctuations (ZPF):

\begin{equation}\label{eq:ZPF}
 S_2^{(eq)}(\omega) =  G \hbar \omega,
\end{equation}
\vskip 0.3cm

\noindent with $G$ the conductance of the system. Experimental
investigations of the shot noise at finite frequency have clearly
shown a constant (voltage independent) noise spectral density for
$\hbar \omega > eV$ in several systems
\cite{Schoelkopf,Portier,GR1}. Although these experiments
 were not able to give an absolute value of the equilibrium noise (because of
intrinsic noise of linear amplifiers used for the measurement),
one has good reasons be believe that ZPF can be observed with this
kind of amplifiers. Indeed, it as been proven in other detection
schemes,  theoretically \cite{Haus,Lesovik} and experimentally
\cite{Koch,Yurke,Deblock,Astafiev}, that ZPF can be detected from
deexcitation of an active detector whereas they cannot be detected
by a passive  detector which is itself effectively in the ground
state.

In view of recent interest in the theory of the full counting
statistics (FCS) of charge transfer, attention has shifted from
the conventional noise (the variance of the current fluctuations)
to the study of the higher cumulants of current fluctuations.
Whereas the discrimination between active and passive detector
seems to be clear for noise spectral density measurement, the
situation is more complex for the measurement of high order
cumulants at finite frequency. Indeed, the issues of detection
scheme are closely related to the problem of ordering quantum
current operators and, if the problem can be solved in a general
way for two operators \cite{Lesovik,Gavish}, measurements of
higher cumulants are pointing out the problem of appropriate
symmetrization of the product of $n$ current operators:

\begin{equation}\label{eq:Sn}
 S_n(\omega) = \langle i(\omega _ 1) \, i( \omega _ 2 ) \, ... \, i( \omega _ n
 )\rangle \; \delta (\omega _ 1 + \omega _ 2 + \, ... \, + \omega _ n)
\end{equation}
\vskip 0.3cm

\noindent It is the goal of this paper to clearly present  the
problem of the third cumulant measurement on a well defined
experimental setup using a linear amplifier as a detector. Until
now, measurements of the third cumulant $S_{3}$ of voltage
fluctuations have been performed at low frequency, \textit{i.e.}
in the classical regime $\hbar \omega < eV,\, k_BT$ where voltage
fluctuations arise from charge transfer process
\cite{Reulet,Reznikov,Leturcq}. We report here the first
measurement of $S_3$ at high frequency, in the quantum regime
$\hbar \omega > eV,\, k_BT$. It raises central questions of the
statistics of quantum noise, in particular:

\vskip 0.5cm

\begin{description}
\item[1.] The electromagnetic environment of the sample has been
proven to strongly influence the measurement, through the possible
modulation of the noise of the sample \cite{Reulet}.  What happens
to this mechanism in the quantum regime?

\item[2.] For $eV < \hbar \omega$, the noise is due to ZPF and keeps its equilibrium value: $S_2= G \hbar
\omega$ with $G$ the conductance of the sample. Therefore, $S_2$
is independent of the bias voltage and no photon is emitted by the
conductor. Is it possible, as suggested by some theories
\cite{Zaikin1,Zaikin2,Hekking}, that $S_3 \neq 0$ in this regime?
\end{description}

In the spirit of these questions, we give theoretical and
experimental answers to the environmental effects showing that
they involve dynamics of the quantum noise. We study the case of a
 tunnel junction, the simplest coherent conductor. Using these
results, we investigate the question of the third cumulant of
quantum noise.

\bigskip

\section{\bf Environmental Effects and Dynamics of Quantum Noise}

We show in this section that the noise dynamics is a central
concept in the understanding of environmental effects on quantum
transport. First, we present a simple approach (in the zero
frequency limit) to calculate the effects of the environment on
noise measurements in terms of the modification of probability
distribution $P(i)$ of current fluctuations. We do not provide a
rigorous calculation, but simple considerations that bear the
essential ingredients of the phenomenon. This allows us to
introduce the concept of noise dynamics and determine the correct
current correlator which describes it at any frequency. Then, we
report the first measurement of the dynamics of quantum noise in a
tunnel junction. We observe that the noise of the tunnel junction
responds in phase with the ac excitation, but its response is not
adiabatic, as obtained in the limit of slow excitation. Our data
are in quantitative agreement with a calculation we have
performed.

\vskip 0.5cm
\begin{figure}[h]
  \begin{center}
     \includegraphics[width=8cm]{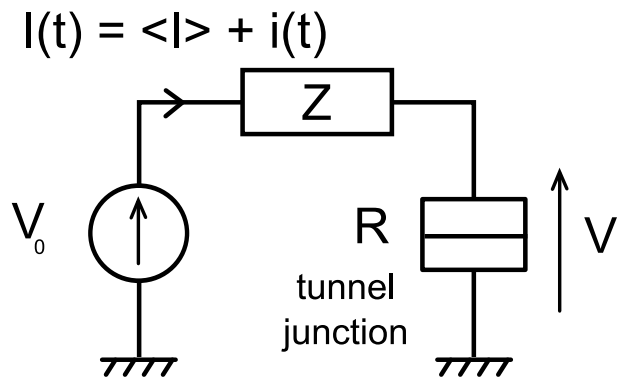}
     \vspace*{0.3cm}
      \caption{Schematics of the experimental setup. Current fluctuations $i(t)=I(t) - \langle I \rangle$
      are measured by an ampmeter with a bandwidth $\Delta f$.
        \label{fig:P(i)}}
  \end{center}
\end{figure}

\vskip 0.5cm

\subsection{Effects of the environment on the probability distribution $P(i)$}

In the zero-frequency limit, high order moments are simply given
by the  probability distribution of the current $P(i)$ calculated
from the current fluctuations measured in a certain bandwidth
$\Delta f$ (see fig.\ref{fig:P(i)}):

\begin{equation}\label{eq:Mn}
 M_n = \int i^n \, P(i) \, di
\end{equation}
\vskip 0.3cm

\noindent The cumulant of order $n$, $S_n$ is then given by a
linear combination of $M_k \, \Delta f ^{k-1}$, with  $k \leq n$
\cite{van}. In practice, it is very hard to perfectly voltage-bias
a sample at any frequency and one has to deal with the non-zero
impedance of the environment $Z$ (see Fig. \ref{fig:P(i)}). If $V$
fluctuates, the probability $P(i)$ is modified. Let us call
$P(i;V)$ the probability distribution of the current fluctuations
around the dc current $I$ when the sample is perfectly biased at
voltage $V$, and $\widetilde{P}(i)$ the probability distribution
in the presence of an environment. $R$ is the resistance of the
sample, taken to be independent of $V$. If the sample is biased by
a voltage $V_0$ through an impedance $Z$, the dc voltage across
the sample is $V =  R_{\parallel}/Z \, V_0$ with  $R_{\parallel}=
RZ/(R + Z)$. The current fluctuations in the sample flowing
through the external impedance induce voltage fluctuations across
the sample, given by:

\begin{equation}\label{eq:dV}
 \delta V(t) = - \int_{-\infty}^{+\infty} Z(\omega)  i(\omega) \,
 e^{i \omega t} \, d \omega
\end{equation}
\vskip 0.3cm

\noindent Consequently, the probability distribution of the
fluctuations is modified. This can be taken into account if the
fluctuations are slow enough that the distribution $P(i)$ follows
the voltage fluctuations. Under this assumption one has:

\begin{equation}\label{eq:dP}
 \widetilde{P}(i)= P(i;V+ \delta V) \simeq P(i,V)+ \delta V \,
 \frac{\partial P}{\partial V} + ...
\end{equation}
\vskip 0.3cm

\noindent supposing that the fluctuations are small ($\delta V \ll
V$). One deduces the moments of the distribution (to first order
in $\delta V=-Zi $) for a frequency independent $Z$:

\begin{equation}\label{eq:dM}
 \widetilde{M}_n \simeq M_n - Z \,
 \frac{\partial M_{n+1}}{\partial V} + ...
\end{equation}
\vskip 0.3cm

\noindent This equation, derived in Ref. \cite{Bertrand_Houches},
shows that environmental correction to the moment of order $n$ is
related to the next moment of the sample perfectly voltage biased.
For $n=1$ we recover the link between noise and Dynamical Coulomb
Blockade through the noise susceptibility (see below) that appears
as $\partial M_2/\partial V$ in the simple picture depicted here
\cite{GR2}. Let us now apply the previous relation to the third
cumulant ($S_n=M_n \, \Delta f^{n-1}$ for $n=2,3$):

\begin{equation}\label{eq:dS3}
 \widetilde{S}_3 \simeq S_3 - 3 Z \,
 \frac{S_{4}}{\partial V} \simeq S_3 - 3 Z \, S_2
 \frac{S_{2}}{\partial V}
\end{equation}
\vskip 0.3cm

\noindent It is a simplified version of the relation derived in
refs. \cite{Nazarov}. The way to understand this formula is the
following: the first term on the right is the intrinsic cumulant;
the second term comes from the sample current fluctuations $i(t)$
inducing voltage fluctuations across itself. These modulate the
sample noise $S_2$ by a quantity $-Zi(t)dS_2/dV$. This modulation
is in phase with the fluctuating current $i(t)$, and gives rise to
a contribution to the third order correlator $\langle
i^3(t)\rangle$. This environmental contribution involves the
impedance of the environment and the \emph{dynamical response} of
the noise which, in the adiabatic limit considered here, is given
by $dS_2/dV$. However, at high enough frequencies, and  in
particular in the quantum regime $\hbar \omega > eV$, this
relation should be modified to include photo-assisted processes.
The notion of dynamical response of the noise is extended in the
following section to the quantum regime in order to subtract
properly the environmental terms in the measurement of the third
cumulant.

\subsection{Dynamics of Quantum Noise in a Tunnel Junction under
ac Excitation}

 In the same way as the complex ac conductance
$G(\omega_0)$ of a system measures the dynamical response of the
average current to a small voltage excitation at frequency $\omega
_0$, we investigate the dynamical response of current fluctuations
$\chi_{\omega_0}(\omega)$, that we name \emph{noise
susceptibility}. It measures the in-phase and out-of-phase
oscillations at frequency $\omega _0$ of the current noise
spectral density $S_2(\omega)$ measured at frequency $\omega$. In
order to introduce the correlator that describes the noise
dynamics, we start with those which describe noise and
photo-assisted noise. Beside the theoretical expressions, we
present the corresponding measurements on a tunnel junction
\cite{Lafe}. It allows to calibrate the experimental setup and
give quantitative comparisons between experiment and theory.

\vskip 0.3cm

\noindent \emph{Noise and photo-assisted noise} \vskip 0.1cm

\begin{figure}
\begin{center}
\hspace*{-1.5cm}\includegraphics[width=10cm]{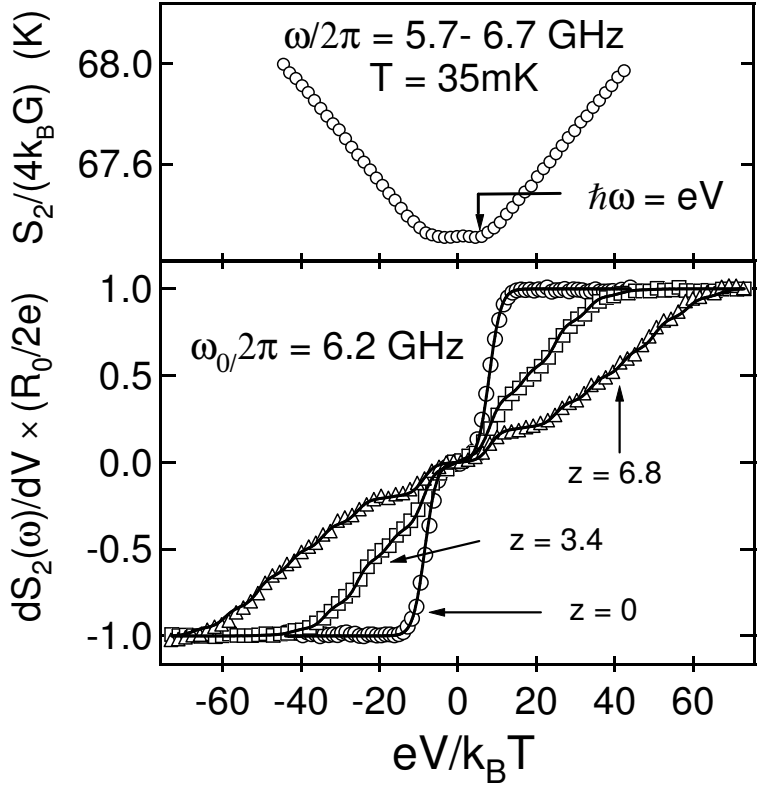}
\vspace*{0.5cm} \caption{Top: Measured noise temperature
$T_N=S_2(\omega)/(4k_BG)$ of the sample plus the amplifier with no
ac excitation. Bottom: measured differential noise spectral
density $dS_2(\omega)/dV$ for various levels of excitation
$z=e\delta V/(\hbar\omega_0)$. $z\neq 0$ corresponds to
photo-assisted noise. Solid lines are fits with Eq.
(\ref{eqSpa}).} \label{fig:s2ds2}
\end{center}
\end{figure}

\begin{figure}
  \begin{center}
     \includegraphics[width=10cm]{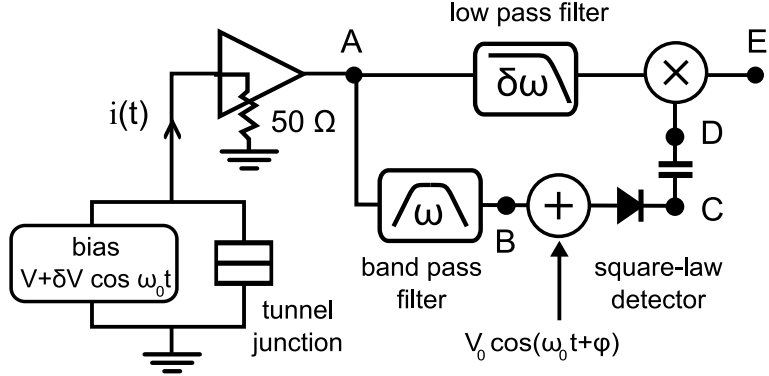}
      \vspace*{0.3cm}
      \caption{ Experimental setup for the measurement of the noise dynamics $X(\omega_0,\omega)$
      and the third cumulant $S_3(\omega, \omega_0 - \omega)$  for $\omega \sim \omega_0$.
       The symbol $\oplus$ represents a combiner, which output is the sum of its two inputs.
       The symbol $\otimes$ represents a multiplier, which output is the product of its two inputs.
       The diode symbol represents a square law detector, which output is proportional to the low frequency part of the square of its
       input.}
        \label{fig:setup}
  \end{center}
\end{figure}

\noindent The spectral density of the current fluctuations at
frequency $\omega$ of a tunnel junction (\textit{i.e.} with no
internal dynamics) biased at a dc voltage $V$ is \cite{Blanter} :
\begin{equation}
S_2(V,\omega)=\frac{S_2^0( \omega_+)+S_2^0 ( \omega_-)}{2},
\label{eqSvsS0}
\end{equation}\vskip 0.3cm

\noindent where $\omega_\pm=\omega\pm eV/\hbar$. $S_2^0(\omega)$
is the Johnson-Nyquist equilibrium noise,
$S_2^0(\omega)=2G\hbar\omega\coth \left(\hbar\omega/(2k_BT)
\right)$ and $G$ is the conductance. At low temperature, the $S_2$
vs. $V$ curve (obtained at point C on Fig. \ref{fig:setup}) has
kinks at $eV=\pm\hbar\omega$, as clearly demonstrated in our
measurement, see Fig. \ref{fig:s2ds2} top. The temperature of the
electrons is obtained  by fitting the data with Eq. \ref{eqSvsS0}.
We obtain $T= 35 \, \mathrm{mK}$, so that $\hbar \omega /k_BT \sim
8.5$. Note that a huge, voltage independent, contribution $T_N
\sim 67 \, \mathrm{K}$ is added to the voltage dependent noise
coming from the sample  which masks the contribution from the ZPF.
When an ac bias voltage $\delta V\cos\omega_0t$ is superimposed on
the dc one, the electrons wavefunctions acquire an extra factor
$\sum_n J_n(z)\exp (in\omega_0t)$ where $J_n$ is the Bessel
function of the first kind and $z=e\delta V/(\hbar\omega_0)$.  The
noise  at frequency $\omega$ is modified by the ac bias, to give:
\begin{equation}
S_2^{pa}(V,\omega)=\sum_{n=-\infty}^{+\infty}J_n^2(z)
S_2(V-n\hbar\omega_0/e,\omega) \label{eqSpa}
\end{equation}
\vskip 0.3cm

\noindent  This effect, called photo-assisted noise, has been
measured for $\omega=0$ \cite{Schoelkopf}. We show below the first
measurement of photo-assisted noise at finite frequency $\omega$.
The multiple steps separated by $eV=\hbar\omega_0$ are well
pronounced and a fit with Eq. \ref{eqSpa} provides the value of
the rf coupling between the excitation line and the sample $\delta
V$ (see Fig. \ref{fig:s2ds2} bottom).

\vskip 0.3cm

\noindent \emph{Noise susceptibility}  \vskip 0.1cm

\noindent Photo-assisted noise corresponds to the noise
$S_2(\omega)$ in the presence of an excitation at frequency
$\omega_0$, obtained by time averaging the square of the current
filtered around $\omega$, as in \cite{Schoelkopf} for $\omega=0$
 and in \cite{GR1} for $\omega\sim\omega_0$. This is similar to the photo-voltaic effect for the dc current. The equivalent of the dynamical
 response of current at arbitrary frequencies $\omega_0$ is the dynamical response of noise at frequency
 $\omega_0$. It involves the beating of
two Fourier components of the current separated by $\pm \omega _0$
expressed by the correlator $\langle i(\omega)
i(\omega_0-\omega)\rangle$ . Using the techniques described in
\cite{Blanter}, we have calculated the correlator that corresponds
to our experimental setup, using the "usual rules" of
symmetrization for a two current correlator and a classical
detector. We find the dynamical response of noise for a tunnel
junction \cite{GR2}:

\begin{equation}\label{eq:X}
X(\omega_0,\omega)  =  \frac{1}{2} \,\sum_n J_n(z)J_{n+1}(z) \,
\left( S_2^0(\omega_+ +n\omega_0)
   -S_2^0(\omega_- -n\omega_0) \right)
\end{equation}

\vskip 0.3cm

\noindent Note the similarity with the expression giving the
photo-assisted noise, Eq. (\ref{eqSpa}). Note however that the sum
in Eq. (\ref{eq:X}) expresses the \emph{interference} of the
processes where $n$ photons are absorbed and $n\pm1$ emitted (or
vice-versa), each absorption / emission process being weighted by
an amplitude $J_n(z)J_{n\pm1}(z)$.

\vskip 0.3cm

\noindent \emph{Experimental setup}  \vskip 0.1cm

\noindent The sample is an Al/Al oxide/Al tunnel junction
identical to that used for noise thermometry \cite{Lafe}. We apply
a 0.1 T perpendicular magnetic field to turn the Al normal. The
junction is mounted on a rf sample holder placed on the mixing
chamber of a dilution refrigerator. The resistance of the sample
$R_0=44.2 \, \Omega$ is close to $50 \, \Omega$ to provide a good
matching to the coaxial cable and avoid reflection of the ac
excitation. The sample is dc voltage biased, ac biased at
$\omega_0/2\pi=6.2$ GHz, and ac coupled to a microwave 0.01-8 GHz
cryogenic amplifier. To preselect the high-frequency component
$i(\omega)$, we use a $5.7-6.7$ GHz band-pass filter (Fig.
\ref{fig:setup}, lower arm). Its beating frequency $\omega$  is
shifted  to low frequency $\delta \omega$ by using a square law
detector and the reference signal $V_0 \cos (\omega_0 t+ \varphi)$
in order to mix it with the low-frequency component $i(\delta
\omega)$. The power detector has an output bandwidth of
$\delta\omega/2\pi\sim 200$ MHz, which limits the frequencies
$\omega$ contributing to the signal:
$|\omega|\in[\omega_0-\delta\omega,\omega_0+\delta\omega]$. The
low frequency part of the current, at frequency $\omega-\omega_0$,
is selected by a $200 \, \mathrm{MHz}$  low pass filter (Fig.
\ref{fig:setup}, upper arm).

\vskip 0.3cm

\noindent \emph{Experimental results}  \vskip 0.1cm

\begin{figure}[h]
  \begin{center}
    \hspace*{-1.5cm} \includegraphics[width=8.5cm]{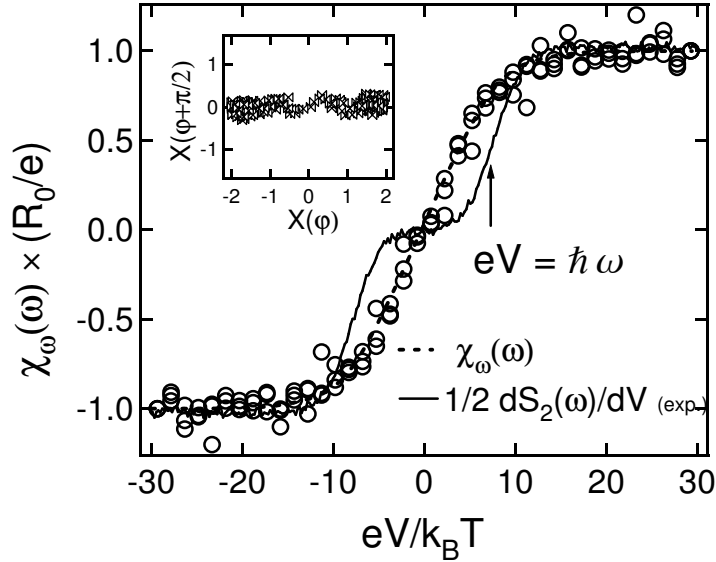}
      \vspace*{0.3cm}
      \caption{ Normalized noise susceptibility $\chi_{\omega}(\omega)$ vs. normalized dc bias.
       Symbols: data for various levels of excitation ($z = 0.85$, $0.6$ and $0.42$). Dotted and dashed lines: fits of $\chi_{\omega}(\omega)$
       (Eq. (\ref{eq:chi0})).
       Solid line: $(1/2)dS_2/dV$ (experimental), as a comparison. \textit{Inset}:  Nyquist representation of $X(\omega_0,\omega)$ for $z=1.7$
       (in arbitrary units). The in-phase and out-of-phase responses are measured by shifting the phase $\varphi$ of the reference signal by $90 ^{\circ}$.
        \label{fig:chi}}
  \end{center}
\end{figure}

\noindent  We could not determine the absolute phase between the
detected signal and the excitation voltage at the sample level.
However we have varied the phase $\varphi$ to measure the two
quadratures of the signal. We have always found that all the
signal can be put on one quadrature only (independent of dc and ac
bias, see inset of Fig. \ref{fig:chi} (b)), in agreement with the
prediction. In the case of as small voltage excitation, we define
the noise susceptibility which is for noise the equivalent of the
ac conductance for current:

\begin{equation}
\chi_{\omega_0}(\omega)=\lim_{\delta V\rightarrow0}
\frac{X(\omega_0,\omega)}{\delta V}
\end{equation}
\vskip 0.3cm

 \noindent $\chi_{\omega_0}(\omega)$ expresses the
effect, to first order in $\delta V$, of a small excitation at
frequency $\omega_0$ to the noise measured at frequency $\omega$.
We show in Fig. \ref{fig:chi} the data for
$X(\omega_0,\omega)/\delta V$ at small injected powers as well as
the theoretical curve for $\chi_{\omega_0}(\omega=\omega_0)$:

\begin{equation}
\chi_{\omega}(\omega)=\chi_{\omega}(0)= \frac{e}{2\hbar\omega} \,
\left( S_2^0(\omega_+)-S_2^0(\omega_-) \right) \label{eq:chi0}
\end{equation}
\vskip 0.3cm

\noindent All the data fall on the same curve, as predicted, and
are very well fitted by the theory. The cross-over occurs now for
$eV\sim\hbar\omega$. $\chi_{\omega}(\omega)$ is clearly different
from the adiabatic response of noise $dS_2(\omega)/dV$ (solid line
in Fig. \ref{fig:chi}). However, in the limit $\delta
V\rightarrow0$ and $\omega_0\rightarrow 0$ (with $z\ll1$), Eq.
(\ref{eq:chi0}) reduces to $\chi_\omega(0)\sim (1/2)(dS_2/dV)$.
The factor $1/2$ comes from the fact that the sum of frequencies,
$\pm(\omega+\omega_0)$ (here $\sim12$ GHz), is not detected in our
setup. This is the central result of our work: the quantum noise
responds in phase but non-adiabatically.

\section{\bf Third Cumulant of  Quantum Noise Fluctuations}

\subsection{Operator odering}

A theoretical framework  to analyze FCS was developed in Ref.
\cite{Levitov} to evaluate any cumulant of the current operator in
the zero-frequency limit. In order to analyze frequency dispersion
of current fluctuations it is necessary to go beyond the usual FCS
theory \cite{Zaikin1,Zaikin2,Hekking}. An essential problem in
these approaches is to know what ordering of current operators
$\hat i$ corresponds to a given detection scheme. This problem is
simpler for $S_2$: the correlator $S_+(\omega)=\langle \hat
i(\omega)\hat i(-\omega)\rangle$ with $\omega>0$ represents what
is measured by a detector that absorbs the photons emitted by the
sample, like a photo-multiplier. The correlator
$S_-(\omega)=\langle \hat i(-\omega)\hat
i(\omega)\rangle=S_+(-\omega)$ represents what the sample absorbs,
and can be detected by a detector in an excited state that decays
by emitting photons into the sample. Finally  a classical detector
cannot separate emission from absorptions, and measures the
symmetrized quantity:

\begin{equation}\label{eq:S2sym}
S_2^{sym.}(\omega)= \frac{  \langle \hat{i}(\omega) \, \hat{i}(-
\omega )\rangle + \langle \hat{i}(- \omega) \, \hat{i}( \omega
)\rangle}{2},
\end{equation}
\vskip 0.3cm

\noindent However, according to Kubo formula
$S_+(\omega)-S_-(\omega)=G\hbar\omega$ , all these contributions
have similar voltage and temperature dependence, at least for a
linear conductor. In contrast, different ordering of three current
operators give rise to very different results. The prediction for
the Keldysh ordering, which is supposed to correspond to a
classical detection, is: $S_3(\omega,\omega')=e^2I$, independent
of frequency even in the quantum regime. As far as we know there
is no clear interpretation of this ordering in terms of absorption
and emission of photons. We give below two detection schemes for
the measurement of $S_3(\omega ,0)$ that may lead to different
results.

\subsection{Measurement of $S_{v^3}(\omega,0)$}

\vskip 0.3cm

\noindent \emph{Experimental setup} \vskip 0.1cm

\noindent We have investigated the third cumulant
$S_{v^3}(\omega,0)$ of the voltage fluctuations of a tunnel
junction in the quantum regime $\hbar \omega >eV$. For technical
reasons (the input impedance of the rf amplifier is fixed at $Z=
50 \, \Omega$), we measured voltage fluctuations $v(t)$ instead of
current fluctuations $i(t)$. Thus, the impedance $R_{\parallel}=
RZ/(R + Z)$ will act as the environment and will affect the
measurement of the third cumulant $S_{3}(\omega,0)$ of the current
fluctuations. We use the same experimental setup and sample as for
the noise dynamics measurement, the only change is that the ac
excitation is switched off: $\delta V = V_0=0$ (see Fig.
\ref{fig:setup}. Thus only the noise of the amplifier can modulate
the noise of the sample. A $5.7 - 6.7 \, \mathrm{GHz}$ band-pass
filter followed by a square law detector allows to mix
high-frequency components $v(\omega) \, v(-\omega- \delta \omega)$
which are multiplied by low-frequency components selected by a
$200 \, \mathrm{MHz}$ low pass filter, we end up with a dc signal
proportional to  $S_{V^3} \propto \langle v(\omega) \, v(-\omega-
\delta \omega) \, v( \delta \omega) \rangle$. The fact that the
same setup is used to detected $S_3$ and $\chi$ is quite
remarkable: it clearly indicates that the environmental effects in
$S_3$ are indeed described by $\chi$ and not by $dS_2/dV$.

\vskip 0.3cm

\begin{figure}[h]
  \begin{center}
     \includegraphics[width=10cm]{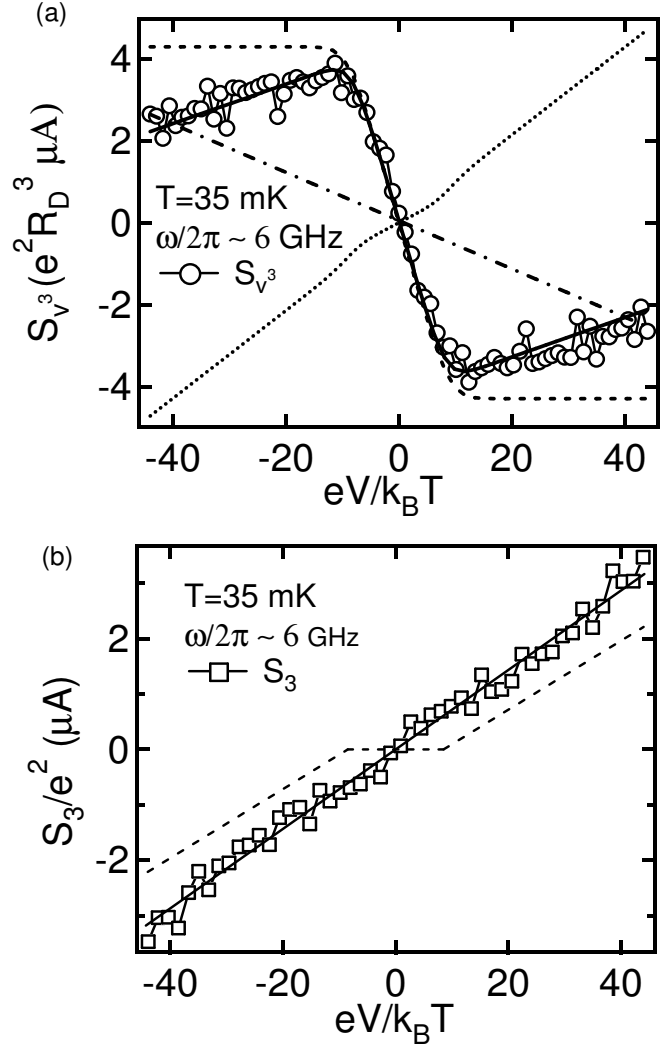}
      \vspace*{0.3cm}
      \caption{(a) Measurement of $S_{v^3}(\omega,0)$ vs. bias voltage $V$ (circles). The solid line corresponds to
the best fit with Eq. (\ref{eq:Sv3}). The dash dotted line
corresponds to the perfect bias voltage contribution and the
dotted lines to the effect of the environment. (b) Measurement of
$S_3(\omega,0)$  vs. bias voltage $V$ (squares). \label{fig:sv3}}
  \end{center}
\end{figure}

\vskip 0.3cm

\noindent \emph{Experimental results} \vskip 0.1cm

\noindent $S_{v^3}$ at $T = 35 \, \mathrm{mK}$ is shown in Fig.
\ref{fig:sv3} (a), these data were averaged for $4$ days. These
results are clearly different from the voltage bias result because
of the environmental contributions. As described before (see
section 1), the noise of the sample is modulated by its own noise
and by the noise of the amplifier $S_{2,N}$, to give rise to an
extra contribution to $S_{v^3}$. By generalizing the expression
(\ref{eq:dS3}), we find, assuming real, frequency independent
impedances to simplify the expression (but we used the full
expression for the fits of the data):

\begin{equation}\label{eq:Sv3}
\begin{array}{ll}
S_{v^3}(\omega,0) = & -R_{\parallel}^3 \, S_3(\omega,0) \, + \,
R_{\parallel}^4  \, \left(S_{2,N}(0)+ S_{2}(0) \right)
\chi_{0}(\omega) \, + \\
&\\
&+ \, R_{\parallel}^4  \, \left(S_{2,N}(\omega)+ S_{2}(\omega)
\right)  \chi_{\omega}(0)  \,+ \, R_{\parallel}^4 \,
\left(S_{2,N}(\omega)+ S_{2}(\omega) \right)
\chi_{\omega}(\omega)\\
\end{array}
\end{equation}
\vskip 0.3cm

\noindent To properly extract the environmental effects, we fit
the data obtained at different temperatures ($35 \, \mathrm{mK}$,
$250 \, \mathrm{mK}$, $500  \, \mathrm{mK}$, $1 \, \mathrm{K}$,
$4.2 \, \mathrm{K}$). The parameters $R_{\parallel}(0)$,
$R_{\parallel}(\omega)$, $S_{2,N}(0)$ and $S_{2,N}(0)$ that
characterize the environment are independent of temperature,
whereas $S_2(V)$ and $\chi(V)$ have temperature dependent shapes.
This allows for a relatively reliable determination of the
environmental contribution. We have performed independent
measurements of these parameters and obtained a reasonable
agreement with the values deduced from the fit. However more
experiments are needed with another, more controlled environment,
to confirm our result. The intrinsic $S_{3}$ in the quantum
regime, obtained after subtraction of the environmental
contributions, is shown in Fig. \ref{fig:sv3} (b). It seems to
confirm the theoretical prediction by \cite{Zaikin2,Hekking}
(solide line), \textit{i.e.} $S_3(\omega,0)=e^2 I$ even for $\hbar
\omega > eV$.

\vskip 0.5cm

\begin{figure}[htbp]
\begin{center}
\includegraphics[width=8cm]{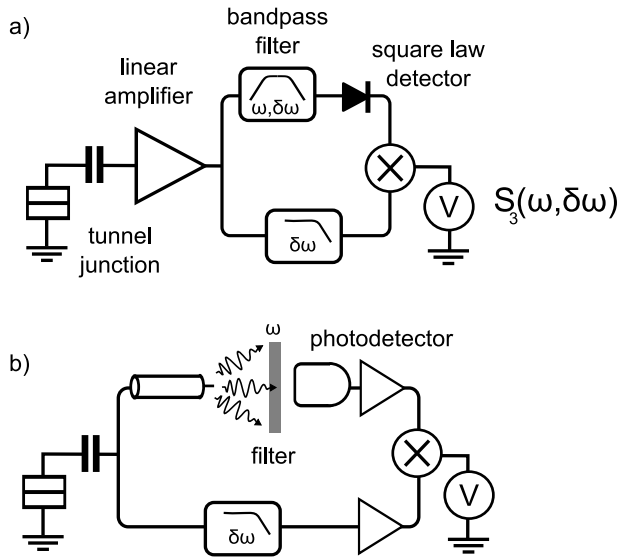}
\end{center}
\vspace*{0.5cm} \caption{(a) Experimental detection scheme. The
symbol $\bigotimes$ represents a multiplier, which output is the
product of its two inputs. The diode symbol represents a square
law detector, which output is proportional to the low frequency
part of the square of its input. $S_3(\omega, \delta \omega
\rightarrow 0)$ is given by the product of the square of high
frequency fluctuations with low frequency fluctuations. (b)
Equivalent detection scheme using a photodetector to measure
square of high frequency fluctuations.} \label{fig:photodetect}
\end{figure}

\vskip 0.5cm

\section{\bf Conclusion}

We have shown the first measurement of the noise susceptibility,
in a tunnel junction in the quantum regime $\hbar \omega \sim
\hbar \omega _0 \gg k_BT $ ($\omega/2\pi \sim 6 \, \mathrm{GHz}$
and $T \sim 35 \, \mathrm{mK}$) \cite{GR1}. We have observed that
the noise responds in phase with the excitation, but not
adiabatically. Our results are in very good, quantitative
agreement with our prediction based on a new current-current
correlator $\chi_{\omega_0}(\omega) \propto \left\langle i(\omega)
i(\omega_0-\omega) \right\rangle$. Using the fact that the
environmental contributions to $S_3$ are driven by $\chi$, we have
been able to extract the intrinsic contribution from a measurement
of $\langle
 v^3\rangle$ on a tunnel junction in the quantum regime. Our
experimental setup is based on a "classical" detection scheme
using linear amplifiers (see Fig. \ref{fig:photodetect} (a)) and
the results are in  agreement with theoretical predictions:
$S_3(\omega, 0)$ remains proportional to the average current and
is frequency independent \cite{Zaikin2,Hekking}. This result
raises the intriguing question of the possibility to measure a
non-zero third cumulant in the quantum regime $\hbar \omega > eV$
whereas the noise $S_2(\omega)$ is the same as at equilibrium, and
given by the zero-point fluctuations.

One can think of another way to measure $S_3(\omega,0)$ with a
photodetector (sensitive to photons \emph{emitted} by the sample),
as depicted in fig. \ref{fig:photodetect} (b). In this case $S_3$
is the result of correlations between the low frequency current
fluctuations and the low frequency fluctuations of the flux of
photons of frequency $\omega$ emitted by the sample. Since no
photon of frequency $\omega$ is emitted for $eV<\hbar\omega$, the
output of the photo-detector is zero and $S_3(\omega,0)=0$. The
expectation of such a measurement is sketched by a dashed line in
fig. \ref{fig:sv3} (b). Note that such a detection scheme has
already been applied on laser diodes \cite{Richardson,Grangier}.

\section*{Acknowledgements}
We are very grateful to L. Spietz for providing us with the sample
that he fabricated at Yale University. We thank M. Aprili, M.
Devoret, P. Grangier, F. Hekking, J.-Y. Prieur, D.E. Prober and I.
Safi for fruitful discussions. This work was supported by
ANR-05-NANO-039-02.

\end{document}